\documentclass[conference]{IEEEtran}
\IEEEoverridecommandlockouts
\usepackage{tabularx}
\usepackage{diagbox}
\usepackage{cite}
\usepackage{amsmath,amssymb,amsfonts}
\usepackage{algorithmic}
\usepackage{graphicx}
\usepackage{textcomp}
\usepackage{xcolor}
\def\BibTeX{{\rm B\kern-.05em{\sc i\kern-.025em b}\kern-.08em
    T\kern-.1667em\lower.7ex\hbox{E}\kern-.125emX}}
\begin{document}

\title{What Are They Doing? Joint Audio-Speech Co-Reasoning}

\author{\IEEEauthorblockN{1\textsuperscript{st} Yingzhi Wang}
\IEEEauthorblockA{\textit{Research Center} \\
\textit{Elm}\\
Riyadh, KSA \\
wangyingzhi666@gmail.com}
\and
\IEEEauthorblockN{2\textsuperscript{nd} Pooneh Mousavi}
\IEEEauthorblockA{\textit{Concordia University} \\
\textit{Mila}\\
Montreal, Canada \\
pooneh.mousavi@concordia.ca}
\and
\IEEEauthorblockN{3\textsuperscript{rd} Artem Ploujnikov}
\IEEEauthorblockA{\textit{Université de Montréal} \\
\textit{Mila}\\
Montreal, Canada \\
artem.ploujnikov@umontreal.ca}
\and
\IEEEauthorblockN{4\textsuperscript{th} Mirco Ravanelli}
\IEEEauthorblockA{\textit{Concordia University} \\
\textit{Mila}\\
Montreal, Canada \\
mirco.ravanelli@concordia.ca}
}

\maketitle

\begin{abstract}
In audio and speech processing, tasks usually focus on either the audio or speech modality, even when both sounds and human speech are present in the same audio clip. Recent Auditory Large Language Models (ALLMs) have made it possible to process audio and speech simultaneously within a single model, leading to further considerations of joint audio-speech tasks.


In this paper, we establish a novel benchmark to investigate how well ALLMs can perform joint audio-speech processing. Specifically, we introduce Joint Audio-Speech Co-Reasoning (JASCO), a novel task that unifies audio and speech processing, strictly requiring co-reasoning across both modalities. We also release a scene-reasoning dataset called \textit{"What Are They Doing"}. Additionally, we provide deeper insights into the models' behaviors by analyzing their dependence on each modality.


\end{abstract}


\section{Introduction}
In the field of deep learning, acoustic processing tasks have long been categorized into two primary modalities: audio tasks and speech tasks. This division is based on the nature of the input data and the ultimate objectives. Audio tasks typically focus on non-speech signals, such as Sound Event Detection \cite{sev}, Music Analysis \cite{music_analysis}, Source Separation \cite{ss}, and Signal Enhancement \cite{se}. These tasks require the model to recognize, classify, and process various non-speech sounds to understand the environment, events, or specific audio characteristics. On the other hand, speech tasks are dedicated to human speech processing, not only understanding semantics such as Automatic Speech Recognition \cite{asr} and Spoken Language Understanding \cite{slu}, but also recognizing paralinguistics such as Speaker Recognition \cite{speaker, speaker_dia} and Speech Emotion Recognition \cite{ser, sed}. Traditional models utilize either the audio modality or the speech modality exclusively, even though both can coexist in the same audio clip.


Recently, with the exploration of speech foundation models \cite{foundation}, multiple tasks were attempted to be unified under the same model. 
Until the rise of Auditory LLMs (ALLMs) \cite{ltu-as, qwen-audio, salmonn, wavllm, speechverse}, various audio or speech tasks could be accomplished within the same model by adjusting the task instructions prompt. 
These models typically use separate encoders to extract audio and speech representations, which are then concatenated and fed into a text-based LLM for reasoning and understanding. 
For example \cite{salmonn} applied a Whisper \cite{whisper} speech encoder and a BEATs audio encoder \cite{beats} to respectively extract speech and audio information, then used a window-level Q-Former \cite{q-former} as the connection module to fuse them and pass the fused embedding through a Vicuna LLM \cite{vicuna}. \cite{wavllm} employed a Whisper encoder as the semantic encoder and a WavLM \cite{wavlm} as the acoustic encoder. The encoded representations were concatenated and input into a LLaMa-2 model \cite{llama2} to follow instructions.


Compared to the unification of audio and speech accomplished at the model architecture level, from the task perspective, almost all the tasks used for training and evaluating ALLMs are still limited to a single modality. This includes traditional audio/speech tasks as well as open-ended question-answering tasks such as Audio/Speech Question Answering (AQA/SQA).
From the training perspective, single-modality tasks cannot allow for the development of a truly intelligent audio-speech LLM that can achieve joint reasoning, and they may also lead to known problems such as task overfitting \cite{salmonn}.
From the evaluation perspective, the single-modality benchmarks \cite{dynamic-superb, airbench, audiobench} cannot effectively reflect the model's joint understanding capacity and cannot determine if the model's reasoning is misled by its preference for a certain modality.


\begin{table}[htbp]
\caption{}{Samples extracted from Open-ASQA (Sample 1-3) and from the proposed \textit{What Are They Doing"} dataset (Sample 4-6). Red and blue are used to represent audio and speech-related information. Audio-oriented answer and speech-oriented answer are provided in our dataset for comparison with the correct answer, while the keywords are also available to indicate the essential actions or details that should be reasoned.}
\begin{center}
\begin{tabular}{|l|}
\hline
\textbf{Sample 1 (OpenASQA dataset)}\\
\textcolor{red}{Audio Description: Sound of gunfire.}\\
\textcolor{blue}{Speech Description: Speech of "Someone's been hit!"}\\
Question: What kind of emotion or mood does the audio clip convey?\\
Answer: The audio clip could potentially convey a sense of chaos,\\
urgency, or danger.\\
\hline
\textbf{Sample 2 (OpenASQA dataset)}\\
\textcolor{red}{Audio Description: Dog, Humain Speech.}\\
\textcolor{blue}{Speech Description: Speech of “Get him going, get him going with it.”}\\
Question: What can be inferred from the audio? \\
Answer: It can be inferred that the speech is addressing the dog in the\\
background.\\

\hline
\textbf{Sample 3 (OpenASQA dataset)}\\
\textcolor{red}{Audio Description: Light Rap Music}\\
\textcolor{blue}{Speech Description: Speech of "Life is like an orange you can peel}\\
\textcolor{blue}{and slice when you taste it..."}\\
Question: What mood does the music convey based on the lyrics and\\
melody?\\
Answer: The music conveys a fun and upbeat mood, with lyrics that\\
talk about enjoying life and being energetic.\\
\hline
\textbf{Sample 4 (\textit{What Are They Doing} dataset)}\\
\textcolor{red}{Audio Description: Sound of seagulls and water lapping against a boat.}\\
\textcolor{blue}{Speech Description: Speech of "Yeah! This one is much bigger!"}\\
Question: What are they possibly Doing\\
\textcolor{red}{Audio-Oriented Answer: They are likely enjoying a day at the beach or}\\
\textcolor{red}{sailing on a boat.}\\
\textcolor{blue}{Speech-Oriented Answer: They are likely comparing the size of objects.}\\
Correct Answer: They are likely fishing on a boat.\\
Correct Answer Keyword(s): [Fishing]\\

\hline
\textbf{Sample 5 (\textit{What Are They Doing} dataset)}\\
\textcolor{red}{Audio Description: Sound of baby crying.}\\
\textcolor{blue}{Speech Description: Speech of "The bottle is too cold."}\\
Question: What are they possibly Doing\\
\textcolor{red}{Audio-Oriented Answer: They are likely comforting a crying baby.}\\
\textcolor{blue}{Speech-Oriented Answer: They are likely holding or touching a bottle}\\
\textcolor{blue}{that feels very cold.}\\
Correct Answer: They are likely feeding a baby with a bottle.\\
Correct Answer Keyword(s): [Feed]\\
\hline
\textbf{Sample 6 (\textit{What Are They Doing} dataset)}\\
\textcolor{red}{Audio Description: Sound of gunfire.}\\
\textcolor{blue}{Speech Description: Speech of "He looks so cool in this uniform."}\\
Q: What are they possibly Doing\\
\textcolor{red}{Audio-Oriented Answer: They are likely engaged in a shooting}\\
\textcolor{red}{activity or in a combat situation.}\\
\textcolor{blue}{Speech-Oriented Answer: They are observing someone wearing a}\\
\textcolor{blue}{uniform and commenting on how cool he looks.}\\
Correct Answer: They are watching a movie or a TV show.\\
Correct Answer Keyword(s): [Movie, TV show]\\
\hline
\end{tabular}
\label{tab1}
\end{center}
\end{table}

Despite its rarity, joint audio-speech tasks have been attempted in certain studies. \cite{salmonn} proposed a Speech Audio Co-reasoning task (SAC), which requires the model to understand a spoken question, find evidence from the background audio events or music, and reason from it to answer the question. However, the nature of SAC is to combine spoken language understanding with audio-based reasoning rather than perform strict co-reasoning. \cite{ltu-as} introduced Joint Audio-Speech Understanding together with a public Open-ASQA dataset. GPT-3.5-Turbo was used to generate joint audio-speech question-answering pairs on AudioSet \cite{audioset} and FMA \cite{fma} datasets. However, these GPT-generated data face some issues that prevent the task from being strictly categorized under a joint audio-speech understanding task.
To better understand its limitations, we present some examples of the Open-ASQA dataset in Table 1. The main issues are as follows:

\begin{itemize}

\item There exists a significant overlap between audio information and speech information, making it difficult to determine whether the model's reasoning is led by one certain modality or both two modalities (Sample 1).

\item There is dominant information in one certain modality that is sufficient for good reasoning, even without using the other modality (In Sample 2, the answer can be directly inferred from the audio even without using speech).

\item The designed answer requires the model to perform merely a concatenation of audio-oriented reasoning and speech-oriented reasoning. In this case, it would be more accurate to say that the task requires the model to understand both audio and speech, rather than achieving joint audio-speech understanding (Sample 3).

\end{itemize}

To give insights into the development of truly intelligent audio-speech LLMs and provide a more profound evaluation of their joint reasoning capabilities, we propose Joint Audio-Speech Co-Reasoning (JASCO), a strict joint audio-speech task that addresses the aforementioned limitations. To support this task, we create and open-source the ``\textit{What Are they Doing}" dataset\footnote{https://github.com/BenoitWang/What\_Are\_They\_Doing}, which contains audio clips with both audio sounds and human speech, along with corresponding open-ended scene reasoning question-answering pairs. Furthermore, we benchmark popular ALLMs on this dataset to assess their joint reasoning ability and determine if they exhibit a preference for a particular modality. 




\begin{figure}[t!]
  \centering
  \includegraphics[width=\linewidth]{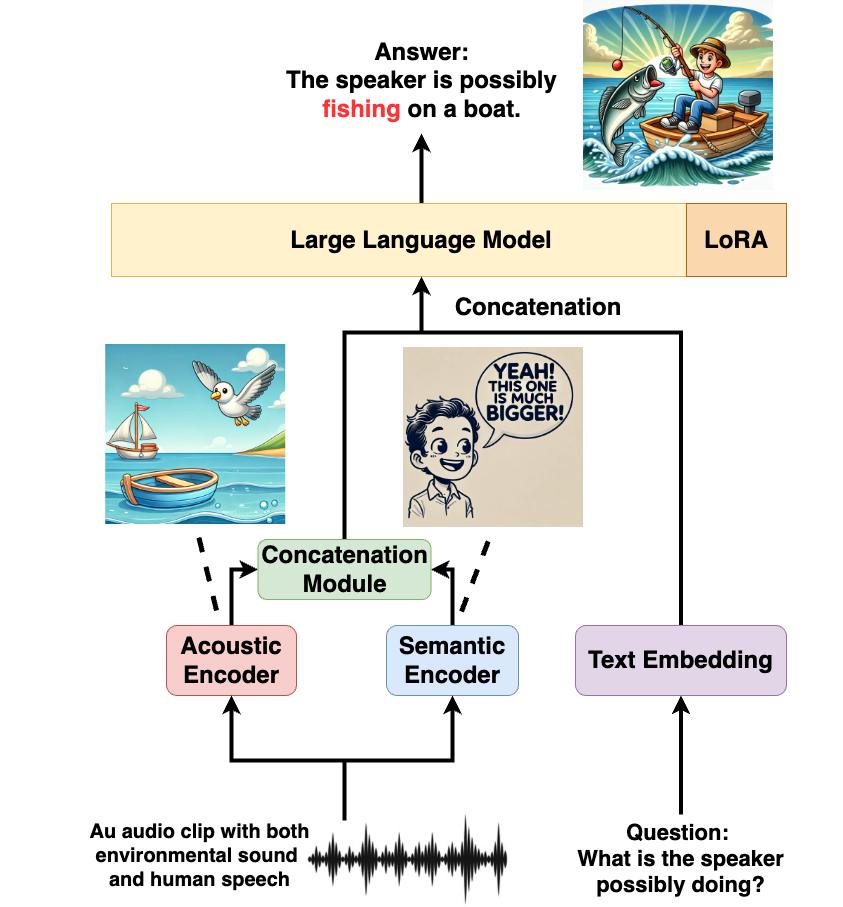}
  \caption{The baseline model architecture for Joint Audio-Speech Co-Reasoning. Dual encoders are used to extract respectively acoustic and semantic information from the input audio clip, which are merged via a concatenation module. The fused embedding is concatenated with the instruction prompt embedding before being passed into an LLM with LoRA adaptors for co-reasoning.}
  \label{fig:4mod}
\end{figure}


\section{Joint Audio-Speech Co-Reasoning}

The Joint Audio-Speech Co-Reasoning task takes an audio clip and a text instruct prompt as input and aims at generating a reasonable response by performing a co-reasoning based on the combination of audio and speech information. In the following, we will detail the specific design considerations for this task, present a baseline model architecture, and explain the metric used for evaluation.

\subsection{Task Design}
We established several constraints to guarantee that the proposed task strictly performs co-reasoning: 

\begin{itemize}

\item \textbf{The audio information and speech information should be irrelevant.} Under this condition, it is intuitive to determine which modality the reasoning depends on.

\item \textbf{Each single modality contains important but non-dominant information.} This constraint forces the model to consider both modalities, as using only one modality alone is insufficient to produce the correct answer.

\item \textbf{Deep co-reasoning is required rather than concatenating information.} This requirement pushes the model to think deeply rather than simply concatenate information.

\end{itemize}
Following these instructions, we propose the \textit{What Are They Doing}  dataset later in Section 3, of which some samples are also shown in Table 1 (Sample 4, 5, and 6). 

\subsection{Baseline Model Architecture}
Thanks to the multi-task processing capability and powerful reasoning performance inherited from LLM \cite{reason}, the ALLM is considered the most suitable baseline model for this task. Therefore, we select the multi-encoder ALLM as the baseline model architecture, as shown in Figure 2.

In the baseline model architecture, pre-trained acoustic encoder (such as BEATs encoder) and semantic encoder (such as Whisper encoder) are used to respectively extract audio and speech information. It should be noted that although paralinguistics (speaker, emotion, etc.) are present in speech modality, they are normally processed with the acoustic encoder. The concatenation module fuses both embeddings, potentially using a connection model such as the Window-level Q-Former used in \cite{salmonn}, or a simple concatenation with adapters on both audio and speech branches as in \cite{wavllm}. The fused audio/speech embedding is then concatenated with the text prompt embedding and fed into a LoRA-adapted \cite{lora} LLM for co-reasoning.

\subsection{Evaluation Metric}
The open-ended nature of the proposed task poses challenges to the evaluation, as traditional metrics may not be suitable. As a result, we use the Model-As-Judge approach \cite{judge, judge2} with a carefully designed prompt to compare the model predictions with the reference correct answers.
The model judge conducts two sub-evaluations for an audio clip:
\begin{itemize}

\item \textbf{Score the prediction.} Evaluate whether the model’s prediction aligns with the reference answer in content, logic, and relevance, using 3 scoring tiers that reflect different levels of the quality of the model's predictions:

\textbf{Score 0}: The prediction is completely misaligned compared to the reference, the model relies on either audio or speech but not both.

\textbf{Score 1}: The inference derived from both audio and speech is logical, generally matching the reference but is missing some important details. 

\textbf{Score 2}: The answer is based on both audio and speech, aligning perfectly with the reference answer and capturing key details.

\item \textbf{Evaluate the modality-dependency.} By exploring the correlation between the model's prediction and the audio/speech information, determine whether the model's reasoning is drawn from both audio and speech or from a single modality alone. The possible results are Audio-Dependent, Speech-Dependent, and Both-Dependent.

\end{itemize}

The complete prompt and examples of the model-judge output on the proposed \textit{What Are They Doing} dataset can also be found in the provided GitHub repository.


\section{\textit{What Are They Doing}  Dataset}

In this section, we introduce the dataset designed to support the joint audio-speech co-reasoning task.

\subsection{Dataset Design}

The design of the dataset involves two critical steps:
\begin{itemize}

\item Following the constraints outlined in Section 2.A, create reasonable triples $\{A, S, B\}$, in which  $A$ refers to audio sound, $S$ refers to spoken text, and $B$ refers to speaker's potential behavior/action.

\item Generate audio clips using the provided audio sound and spoken text.

\end{itemize}

An attempt was initially made to generate the triples using GPT-4 \cite{gpt4}. Although we provided detailed prompts emphasizing the three constraints, the samples generated by GPT-4 did not consistently meet the desired quality. Therefore, we opted to manually construct the triples to ensure they fulfill all three conditions. Then, to assess the quality of these manually created triples, we employed GPT-4 as an annotator. While generating triples posed a huge challenge for GPT-4, evaluating the answer based on given audio and speech information proved much easier. The quality assessment of a triple using GPT-4 involves the following three checks:

\begin{itemize}

\item Verify if $B$ can be inferred given both $A$ and $S$.

\item Verify if $B$ is the most likely answer by posing the question multiple times and with various instructions.

\item Inputting only $A$/$S$ leads GPT-4 to an audio-oriented/speech-oriented answer, verify if they differ completely from the correct answer.

\end{itemize}

If any step does not pass the quality check, minor adjustments should be made to the triple until it meets all the checks.

\subsection{Audio Preparation}

Given the strict constraints of the dataset design, finding pre-existing natural audio clips is almost impossible. Consequently, we decided to manually synthesize the audio. First, we selected the most suitable audio clips based on the designed audio sound from Freesound \cite{freesound}. Then, we used a multi-speaker and multi-emotion TTS model \footnote{https://typecast.ai/} to synthesize speech based on the designed spoken text. Finally, both audio clips were resampled to 16 kHz and merged into a mono-channel audio clip. It is important to note that the speech information covers not only the spoken text but also paralinguistics such as the speaker’s gender, age, and emotion. To make the audio more natural and the inference more reasonable, these features were manually adjusted during the speech generation.

Due to the complexity of the task and the significant manual effort required, we were able to produce 80 high-quality co-reasoning samples. While this quantity is insufficient for model training, it can serve as a unique test set for ALLM evaluation to demonstrate their potential abilities and behaviors. 


\begin{table}[t]
\caption{}{The Best-Mean$\uparrow$ (between 0 and 2) for four popular ALLMs. Three LLM judges were utilized for evaluation and their average scores were computed.}
\begin{center}
\resizebox{\columnwidth}{!}{
\begin{tabular}{|c|c|c|c|c|}
\hline
\diagbox{LLM-Judge}{ALLM} & WavLLM & LTU-AS & SALMONN & Qwen2-Audio \\
\hline
Llama-3.1-70B-Instruct & 0.49 & 1.04 & 1.21 & \textbf{1.38}\\
Qwen2.5-72B-Instruct & 0.33 & 1.19 & 1.29 & \textbf{1.43} \\
Mistral-123B-Instruct & 0.13 & 0.74 & 0.80 & \textbf{0.88} \\
\hline
Average & 0.31 & 0.99 & 1.10 & \textbf{1.23} \\
\hline
\end{tabular}
}
\label{tab1}
\end{center}
\end{table}

\section{Experiments}
\subsection{Experimental Setup}
We benchmarked four popular ALLMs on the proposed \textit{What Are They Doing} dataset:
WavLLM-7B\footnote{https://github.com/microsoft/SpeechT5/tree/main/WavLLM}, LTU-AS-7B\footnote{https://github.com/YuanGongND/ltu/tree/main/src/ltu\_as}, SALMONN-7B\footnote{https://huggingface.co/tsinghua-ee/SALMONN-7B/tree/main}, Qwen2-Audio-Instruct-7B\footnote{https://huggingface.co/Qwen/Qwen2-Audio-7B-Instruct} \cite{qwen2-audio}.
It should be mentioned that almost all the training data of WavLLM are speech-related, so it can be considered a pure speech-LLM and is used as a comparison to other models. For each sample, we tested with 8 different instruction prompts to reduce the bias of the evaluation \cite{promptbench}.
In our experiments, we found that the models did not always make a prediction on the speaker's action. To reduce the difficulty for the models, we included step-by-step thinking \cite{cot} instructions by asking them to first detect the audio sounds and transcribe the spoken texts. An example can be: \textit{"What sound can you hear? What does the speaker say? Guess what activity the speakers are engaged in using both the audio sound and the spoken text."}. 


We scored the ALLMs' predictions using the Model-As-Judge method mentioned in section 2.C. In order to enhance the evaluation's reliability, we selected three open-sourced LLMs as judges and calculated their average. The selected judges are Llama-3.1-70B-Instruct \footnote{https://huggingface.co/meta-llama/Meta-Llama-3.1-70B-Instruct}, Qwen2.5-72B-Instruct\footnote{https://huggingface.co/Qwen/Qwen2.5-72B-Instruct} and Mistral-123B-Instruct\footnote{https://huggingface.co/mistralai/Mistral-Large-Instruct-2411}.


\subsection{Results and Analysis}
In order to reflect the upper limit of the ALLMs' capabilities, for each sample, we selected only the best score among the 8 predictions for computing the average across the entire dataset, which is referred to as \textit{Best-Mean}. For each ALLM, we ran three LLM judges and calculated the average \textit{Best-Mean} among the judges.
As can be seen from Table 2, Qwen2-Audio achieved the highest average \textit{Best-Mean} of 1.23, followed by SALMONN at 1.10 and LTU-AS at 0.99. In contrast, WavLLM only obtained a score of 0.31, suggesting that training with only speech leads to poor performance on the proposed benchmark, which also confirms the validity of the proposed dataset. Besides, all the models exhibit a substantial gap below the highest score 2, indicating their limited co-reasoning power under the proposed JASCO task.

\begin{figure}[t!]
  \centering
  \includegraphics[width=\linewidth]{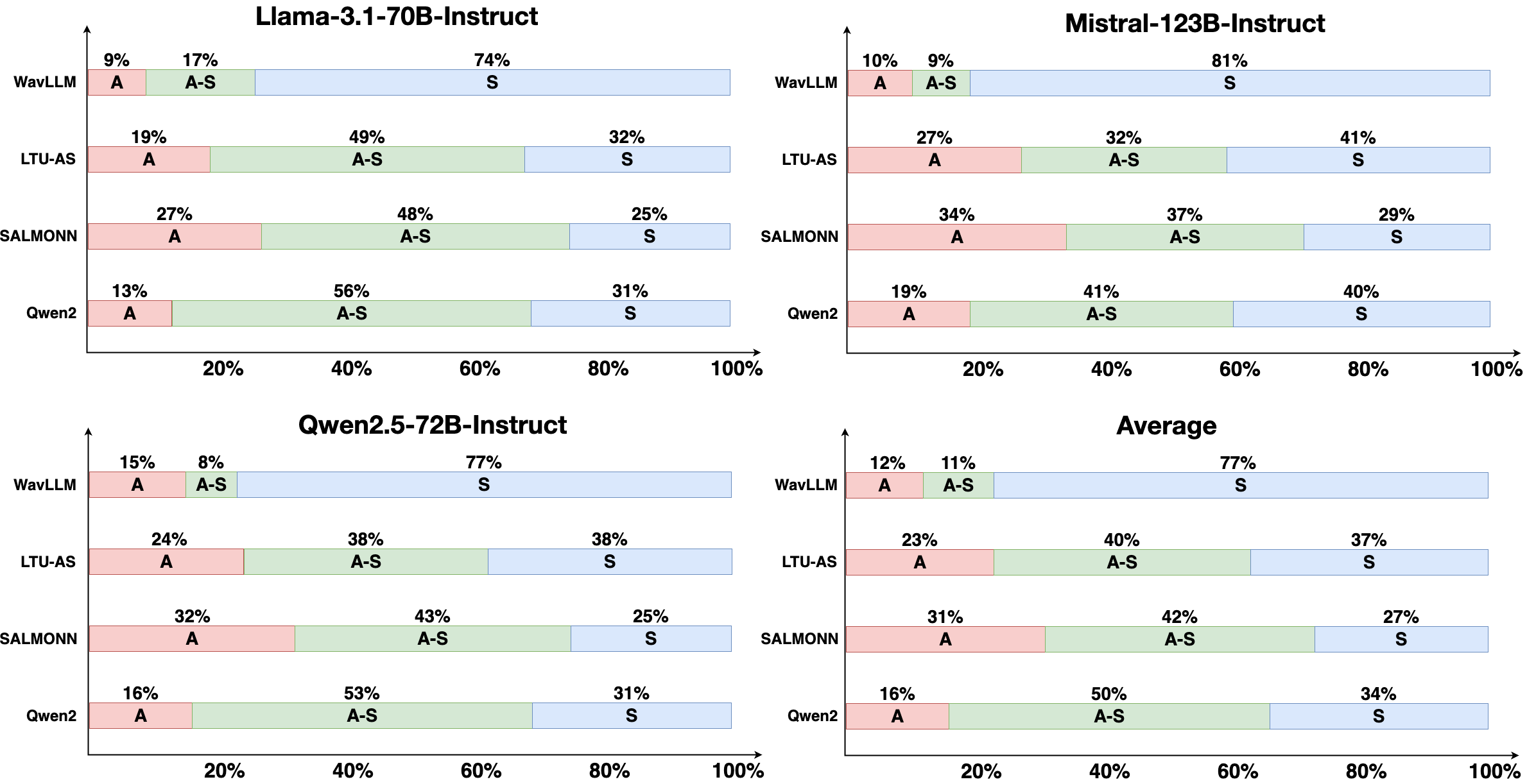}
  \caption{Modality-dependence results of 4 popular ALLMs. The four sub-figures represent evaluations from three LLM judges and their average. Red, blue, and green are used to represent the proportion of audio-dependent (A), speech-dependent (S), and both-dependent (A-S) responses.}
  \label{fig:4mod}
\end{figure}

In Figure 2, we present the results of the ALLMs' modality-dependence. The calculation is based on all the responses obtained using all 8 instruction prompts across all samples. Responses in which the ALLMs did not provide predictions regarding the speaker's actions were excluded from this analysis. Similarly, we report the results from the three judges as well as their average.
As the results indicate, the Qwen2-Audio still leads with an average both-dependent percentage of 50\%, followed by Salmonn at 42\% and LTU-AS at 40\%. In contrast, WavLLM records only 11\% but has the highest average speech-dependence at 77\%, which highlights its nature as a speech LLM. It can be seen from the results that the current ALLMs still lack a consistent practice of joint reasoning. We further discover that despite the equal importance of audio and speech in the JASCO task, both Qwen2 and LTU-AS lean more toward being speech-LLMs than audio-LLMs with always higher speech-dependent percentage than audio-dependent percentage. Unlike the others, Salmonn demonstrates a more balanced dependence on audio and speech. 
By presenting this benchmark, we aim to  provide diagnostic insights into the behavior of existing ALLMs.



\section{Conclusion}
In this work, we built a joint audio-speech benchmark for ALLMs and revealed their modality-dependence. Specifically, we propose the Joint Audio-Speech Co-Reasoning task and the \textit{What Are They Doing} dataset. Through our work, we hope to provide deeper insights into the behavior of existing ALLMs and lay the groundwork for the future development of truly intelligent audio-speech LLMs.

\bibliographystyle{IEEEbib}
\bibliography{strings,refs}

\end{document}